\documentclass[acmsmall, screen, nonacm]{acmart}

\AtBeginDocument{%
  }

\usepackage[rgb,dvipsnames]{xcolor}
\usepackage{soul}
\usepackage{hyperref}
\usepackage[frozencache]{minted}

\usepackage{subcaption}
\usepackage{listings}
\DeclareCaptionSubType*[alph]{listing}
\usepackage[most]{tcolorbox}
\usepackage{tabularx}
\usepackage[ruled,vlined,linesnumbered]{algorithm2e}
\usepackage{subfiles} %

\begin{document}

\title{ProjAgent: Procedural Similarity Retrieval for Repository-Level Code Generation}
\author{Qihong Chen}
\email{chenqh@uci.edu}
\orcid{0009-0000-4844-4717}
\affiliation{%
  \institution{University of California, Irvine}
  \city{Irvine}
  \country{USA}}

\author{Aaron Imani}
\orcid{0000-0001-7183-5468}
\correspondingauthor
\affiliation{%
  \institution{University of California, Irvine}
  \city{Irvine}
  \country{USA}}
\email{aaron.imani@uci.edu}

\author{Iftekhar Ahmed}
\orcid{0000-0001-8221-5352}
\affiliation{%
  \institution{University of California, Irvine}
  \city{Irvine}
  \country{USA}}
\email{iftekha@uci.edu}

\renewcommand{\shortauthors}{Chen et al.}

\begin{abstract}
Repository-level code generation requires implementing target functions while accounting for complex cross-file dependencies and project-specific conventions. Existing retrieval methods predominantly rely on lexical, structural, or semantic similarity, often overlooking repository functions that implement similar procedural logic despite differing in identifiers or application domains. We propose ProjAgent, a repository-level code generation system that introduces procedural similarity as an explicit retrieval signal. ProjAgent decomposes the target function into intermediate reasoning steps and employs an agentic workflow to retrieve repository functions that exhibit similar procedural behavior at each step. The retrieved procedural context is integrated with conventional semantic retrieval to construct a richer repository context for code generation. ProjAgent further incorporates a conservative static-analysis feedback loop that iteratively repairs generated code using compiler and static-analysis feedback. Evaluated on REPOCOD, ProjAgent achieves 41.14\% Pass@1, outperforming existing retrieval-based baselines. These results demonstrate that procedural similarity is an effective and previously unexplored retrieval dimension for repository-level code generation.

\end{abstract}

\maketitle

\section{Introduction}
\label{section:intro}

Large Language Models (LLMs) have demonstrated strong performance across a wide range of Software Engineering (SE) tasks. From automated bug fixing to test case generation, LLM-based approaches have achieved promising results on established benchmarks such as SWE-bench~\cite{jimenez2024swe} and BigCodeBench~\cite{zhuo2025bigcodebench}. These advances have accelerated the adoption of LLMs in software development workflows, supporting activities including code writing, code review, and software maintenance. Among these applications, code generation has attracted particular attention due to its broad practical utility and increasing adoption in real-world development~\cite{ziegler2022productivity,shah2025students}.

Despite these advances, repository-level code generation remains substantially more challenging than standalone code generation~\cite{chen2021evaluating,austin2021program,wang2025can}. Unlike standalone benchmarks, real-world software development rarely involves implementing isolated functions. Instead, developers work within repositories where functions depend on utilities, type definitions, APIs, and project-specific conventions distributed across multiple files~\cite{li2024deveval,zhang2023repocoder}. When relevant repository context is unavailable, LLMs frequently hallucinate APIs, invoke nonexistent functions, or generate implementations that violate project conventions, resulting in incorrect or non-executable code~\cite{le2026not}. Prior work has shown that removing repository context consistently degrades code generation performance across models~\cite{li2024deveval}. Consequently, repository-level code generation depends not only on the generation capability of an LLM, but also on its ability to retrieve repository context that is useful for solving the target programming task.

Existing context retrieval methods for repository-level code generation, such as BM25 and dense embedding search, rely primarily on lexical or semantic similarity~\cite{zhang2023repocoder,gu2025retrieve,robertson2009probabilistic}. These methods were originally developed for code search, where retrieving surface-level similar examples is often sufficient~\cite{le2026not}. In repository-level code generation, however, critical context for the target function (i.e., the function to be generated) may come from helper functions that it depends on, even when those functions differ substantially in naming, data types, or domain vocabulary. Useful context may also come from functions that implement similar logical procedures without sharing a direct dependency relationship. Because such functions often exhibit little lexical or semantic similarity to the target function, they are likely to be overlooked by existing retrieval methods. We refer to functions whose steps share the same underlying reasoning patterns as the target function's steps as a procedurally similar context.

Figure~\ref{lst:retrieval_example} illustrates this challenge. The coding problem \textit{BlackBody.evaluate} contains a step that checks whether the inputs x and temperature are \textit{astropy.units.Quantity} instances and raises ValueError if the temperature is negative. The retrieved context step is from the method \textit{FLRW.m\_nu}, which checks that the input value has the correct shape and is non-negative, and raises a ValueError if either condition fails. The coding problem is in \textit{astropy/modeling/physical\_models.py}, and the retrieved context is in \textit{astropy/cosmology/flrw/base.py}. Although they are located in different modules and serve different purposes, they share a common computational pattern: validating inputs for shape, type, and value constraints. Existing retrieval signals, such as BM25, dense embedding, or data flow analysis, fail to retrieve this context step for this coding problem step because the two steps share no lexical overlap (BM25 similarity of 0.38), low semantic surface similarity (embedding similarity of 0.59), and no direct call dependency between the two modules. As a result, surface similarity-based retrieval can miss critical dependencies while retrieving context that appears relevant but serves a different purpose, ultimately degrading generation quality \cite{le2026not,gu2025retrieve}.

This observation suggests that repository-level retrieval should consider not only lexical and semantic similarity but also \textit{procedural similarity}, the extent to which two functions implement similar computational procedures regardless of naming or domain-specific vocabulary. Procedurally similar functions may arise because one function serves as a building block for another or because both independently implement comparable implementation patterns, such as input validation, state transformation, or unit conversion. Identifying such relationships requires representations that capture implementation behavior beyond surface-level textual similarity. Recent work has shown that intermediate LLM representations encode rich semantic and functional properties of source code~\cite{hu2025harp}, motivating their use for identifying procedurally related functions.

\begin{listing*}[!ht]
\centering
\begin{subcaptiongroup}
\begin{subcaptionblock}{0.45\textwidth}
\centering
\begin{minted}[fontsize=\scriptsize, linenos, breaklines, style=vs, frame=leftline]{python}
def evaluate(self, x, temperature, scale):
    # Check if the input temperature is a valid type (astropy.units.Quantity)
    if not isinstance(temperature, u.Quantity):
        in_temp = u.Quantity(temperature, u.K)
    else:
        in_temp = temperature

    # Check if the input x is a valid type (astropy.units.Quantity)
    if not isinstance(x, u.Quantity):
        in_x = u.Quantity(x, self.input_units["x"])
    else:
        in_x = x

    # Check if temperature is not zero
    if np.any(temp < 0):
        raise ValueError(f"Temperature should be positive: {temp}")
    ...
\end{minted}
\caption{Coding Problem Step: \texttt{BlackBody.evaluate} (step\_0) from \texttt{physical\_models.py}}
\label{lst:example_query}
\end{subcaptionblock}
\hfill
\begin{subcaptionblock}{0.48\textwidth}
\centering
\begin{minted}[fontsize=\scriptsize, linenos, breaklines, style=vs, frame=leftline]{python}
@m_nu.validator
def m_nu(self, param, value):
    if (nneutrinos := floor(self.Neff)) == 0 or self.Tcmb0.value == 0:
        return None  # None, regardless of input

    value = _validate_with_unit(self, param, value)

    # Check values and data shapes and raise ValueError if the shape is incorrect
    if value.shape not in ((), (nneutrinos,)):
        raise ValueError(
            "unexpected number of neutrino masses "
            f"expected {nneutrinos}, got {len(value)}."
        )
    elif np.any(value.value < 0): # check if any mass is negative
        raise ValueError("invalid (negative) neutrino mass encountered.")
    ...
\end{minted}
\caption{Retrieved Context Step: \texttt{FLRW.m\_nu} (step\_2) from \texttt{flrw/base.py}}
\label{lst:example_context}
\end{subcaptionblock}
\end{subcaptiongroup}
\caption{Example of procedurally similar steps. Both steps implement the same guard-clause pattern, validating numerical inputs against domain-specific constraints and raising \texttt{ValueError} on violation, yet share low lexical overlap between their descriptions (``temperature/scale/Quantity'' vs.\ ``neutrino mass/shape/negative'').}
\label{lst:retrieval_example}
\end{listing*}

To leverage this insight, we propose \textsc{ProjAgent}, a repository-level code generation system that combines procedural, lexical, and semantic retrieval signals to construct richer repository context. Rather than relying solely on representation similarity, ProjAgent employs an agentic retrieval workflow that first identifies and validates a small set of procedurally related functions, then expands this set to retrieve additional context using hidden-state projection similarity. Because successful repository-level code generation requires both procedural guidance and repository-specific knowledge, ProjAgent complements procedural retrieval with conventional lexical and semantic retrieval to capture project APIs, variables, and structural dependencies~\cite{zhang2023repocoder,li2024deveval}. Finally, ProjAgent incorporates a static-analysis feedback loop that iteratively refines generated code to improve correctness. We evaluate ProjAgent through the following research questions:

\begin{itemize}
\item \textbf{RQ1:} To what extent does ProjAgent improve repository-level code generation performance compared to existing retrieval-augmented generation baselines?

\item \textbf{RQ2:} How effectively does projection similarity identify procedurally related context for repository-level code generation?

\item \textbf{RQ3:} What is the contribution of each retrieval component to ProjAgent's overall performance?

\end{itemize}

Our contributions are as follows:
\begin{itemize}

\item We introduce \textit{procedural similarity} as a retrieval dimension for repository-level code generation, complementing existing lexical and semantic retrieval signals.

\item We leverage LLM hidden-state projections to represent procedural similarity, enabling the retrieval of procedurally related context across naming and domain boundaries.

\item We propose \textsc{ProjAgent}, a repository-level code generation system that uses an agentic workflow to identify procedural context and combines it with lexical and semantic retrieval for broader context coverage.

\item We conduct a comprehensive evaluation of \textsc{ProjAgent} on repository-level code-generation benchmarks, comparing it against existing retrieval-augmented generation baselines and analyzing the contributions of its retrieval components.

\end{itemize}

The rest of the paper is structured as follows. Section \ref{section:related_work} reviews related work. Section \ref{section:background} introduces the technique for extracting LLM's reasoning. Section \ref{section:method} describes the details of the ProjAgent system. Section \ref{section:exp_setup} presents the experimental setup for answering each research question. Section \ref{section:result} reports the results, and Section \ref{section:ttv} discusses the threat to validity.

\section{Related Work}
\label{section:related_work}

\noindent\textbf{Repository-Level Code Generation and Context Retrieval:}
Repository-level code generation extends traditional code generation by requiring models to generate code that is consistent with cross-file dependencies, project-specific APIs, and repository conventions. Benchmarks such as DevEval~\cite{li2024deveval}, RepoExec~\cite{le2024repoexec}, and REPOCOD~\cite{liang2025can} have highlighted the importance of retrieving relevant repository context for accurate code generation. Consequently, a large body of work has focused on improving repository context retrieval.

Early retrieval-augmented approaches primarily relied on lexical and semantic similarity. RepoCoder~\cite{zhang2023repocoder} introduced an iterative retrieve-then-generate framework based on BM25 and dense retrieval, establishing retrieval-augmented generation as a standard paradigm for repository-level code generation. Subsequent work explored richer repository context beyond textual similarity. CatCoder~\cite{pan2024enhancing} emphasized type-related information, including API definitions and class hierarchies, while RepoExec~\cite{le2024repoexec} demonstrated that carefully selected context can outperform providing the entire repository. More recent methods have incorporated structural information to improve retrieval. DraCo~\cite{cheng2024dataflow} and GraphCoder~\cite{liu2024graphcoder} leveraged program dependency structures to capture cross-file relationships, whereas RepoHyper~\cite{phan2025repohyper} and CodeRAG~\cite{zhang2025coderag} combined multiple complementary retrieval signals, including sparse retrieval, dense embeddings, graph representations, and dataflow analysis.

Several recent approaches have also moved beyond retrieving entire functions. AllianceCoder~\cite{gu2025retrieve} represents repository functions using natural-language descriptions and retrieves context by matching implementation steps to these descriptions. Hydra~\cite{le2026not} further demonstrated that retrieval based primarily on surface similarity can introduce misleading context that degrades generation quality. These findings suggest that effective repository-level code generation depends not only on retrieving semantically relevant code but also on retrieving context that is useful for implementing the target function.

Our work extends this line of research by introducing \textit{procedural similarity} as a complementary retrieval signal. Unlike existing approaches that primarily rely on lexical, semantic, structural, or functional similarity, ProjAgent retrieves functions that implement similar computational procedures, enabling the discovery of useful context even when functions differ substantially in naming, APIs, implementation details, or application domains.

\noindent\textbf{Agentic Workflows for Repository-Level Code Generation:}
Recent work has increasingly adopted agentic workflows to improve repository-level code generation by allowing LLMs to actively explore repositories, retrieve context, and iteratively refine generated code. CodeAgent~\cite{zhang2024codeagent} introduced an autonomous framework equipped with repository exploration tools, including document retrieval and symbol search. TENET~\cite{hu2025tenet} incorporated repository navigation into a test-driven development loop, enabling agents to iteratively retrieve context and repair generated code using execution feedback. SpecAgent~\cite{ma2025specagent} further decomposed repository exploration across specialized agents responsible for context retrieval, planning, and implementation.

Although these systems demonstrate the benefits of agentic workflows for repository understanding and iterative code generation, their retrieval strategies primarily rely on repository navigation, structural dependencies, or surface-level relevance. In contrast, ProjAgent employs an agentic workflow to identify and validate procedurally similar functions before expanding retrieval using hidden-state projection similarity. Rather than introducing another repository exploration strategy, our work introduces a new retrieval signal that complements existing lexical and semantic retrieval methods within an agentic generation framework.

\section{Background}
\label{section:background}

Measuring procedural similarity between functional steps requires representations that capture implementation behavior beyond surface-level lexical similarity. To obtain such representations, we build on the reasoning-subspace projection technique proposed by Hu et al.~\cite{hu2025harp}. We adopt this approach because it isolates hidden-state components associated with reasoning-related behavior, while suppressing linguistic variation. Prior studies have shown that LLM hidden states encode rich semantic and structural information about source code~\cite{zou2023representation,li2022emergent}, making them suitable representations for downstream software engineering tasks.

Hu et al.~\cite{hu2025harp} observed that the hidden-state space of an LLM can be approximately decomposed into a semantic subspace and a reasoning-related subspace. The semantic subspace encodes the linguistic expression, and the reasoning subspace encodes the LLM's internal reasoning process. To identify these subspaces, they perform Singular Value Decomposition (SVD)~\cite{golub2013matrix} on the parameter matrix of the unembedding layer $W_{unemb} = U\Sigma V^\top$,
which maps hidden states to output-token logits. Since the unembedding layer encodes both semantic and reasoning information, the right singular vectors (columns of V) corresponding to the dominant singular values span the semantic subspace, while the remaining vectors span the reasoning subspace $\mathcal{S}_{Reasoning}$. The projection of the hidden state $h_l$ onto the reasoning subspace is computed as: 
\begin{equation}
    \text{proj}_R(h_l) = V_R^\top \cdot h_l
\end{equation}
where $h_l$ is the hidden state at the last layer prior to the final layer normalization, extracted from the response tokens. The resulting representation emphasizes reasoning-related information while suppressing surface-level linguistic variation. We reuse the subspace construction settings from Hu et al. \cite{hu2025harp}, except for the energy threshold, which we increase from 0.95 to 0.98. In the HARP paper, Hu et al. \cite{hu2025harp} reported that the reasoning subspace should occupy approximately 5\% of the hidden state dimension, calculated as $\frac{d - k}{d} \times 100\%$. Here, d is the model's hidden-state dimensionality (model-dependent), and k is determined by the energy threshold. Thus, the goal is to adjust the energy threshold so that the resulting k could yield approximately 5\% of the hidden state dimension for our model. We empirically tested different energy threshold values and found that a threshold of 0.98 could yield 5\% of the hidden state dimension.

While Hu et al.~\cite{hu2025harp} use reasoning-subspace projections as features for hallucination detection, we use them to represent the underlying logic of functional steps and measure procedural similarity through cosine similarity between projected representations. However, we found that the cosine similarity between any pair of projections remained uniformly high even for unrelated functional steps. This behavior is consistent with the anisotropy phenomenon observed in neural representations~\cite{ethayarajh2019contextual,timkey2021all}, where vectors are dominated by a common direction that is largely independent of their content.

To mitigate this issue, we introduce a Principal Component Analysis (PCA) \cite{pearson1901liii} based debiasing step that is not present in the original method of Hu et al.~\cite{hu2025harp}. PCA identifies the dominant direction within a set of representations by computing the mean $\mu$ and the first principal component $pc_1$, which together summarize the common variation shared across all representations. The dominant direction in the presentations can then be removed by subtracting both from each representation. Specifically, we computed $\mu$ and $PC_1$ from a representative set of projected representations as:
\begin{equation}
    \mu = \frac{1}{N}\sum_{i=1}^{N} \text{proj}_R(h_l), \quad \text{PC}_1 = \text{PCA}_1\left(\{\text{proj}_R(h_l)\}_{i=1}^{N}\right)
\end{equation}

To debias each projected representation, we first mean-center it by subtracting $\mu$ and then remove its component along the first principal component $pc1$:
\begin{equation}
    \tilde{h} = (\text{proj}_R(h_l) - \mu) - \left((\text{proj}_R(h_l) - \mu)^\top \cdot \text{PC}_1\right) * \text{PC}_1
    \label{eq:debias}
\end{equation}
Removing the dominant shared direction reduces anisotropy and substantially improves the discriminative ability of cosine similarity when comparing procedural representations.

\section{ProjAgent Pipeline}
\label{section:method}

\begin{figure}[t]
  \centering
  \includegraphics[width=0.85\textwidth]{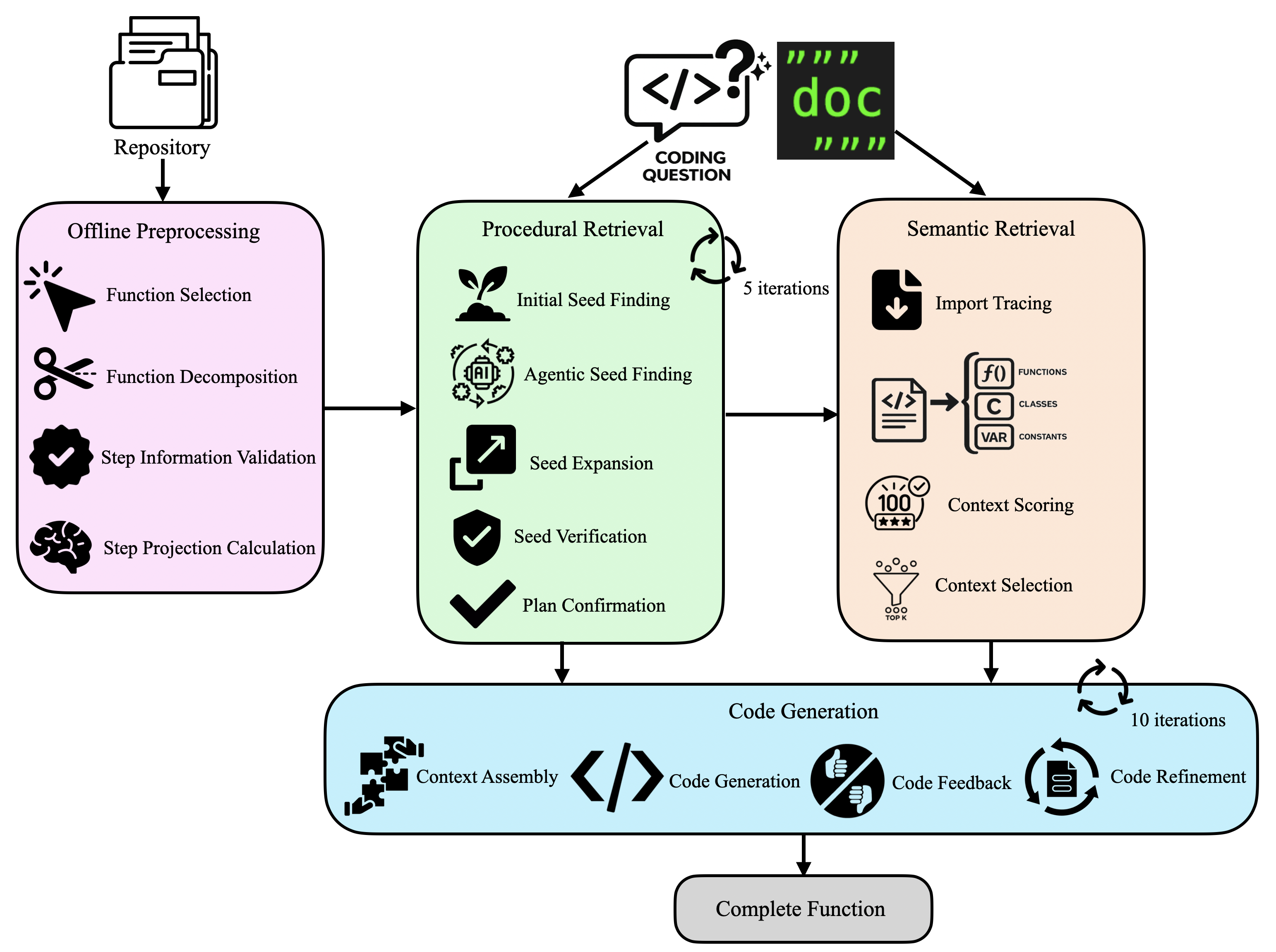}
  \caption{Overview of the ProjAgent pipeline. Stage 1 preprocesses the repository. Stage 2a and Stage 2b retrieve procedural and semantic contexts. Stage 3 generates code using the retrieved contexts and static analysis feedback.}
  \Description[Overview of the ProjAgent pipeline]{Stage 1 preprocesses the repository. Stage 2a and Stage 2b retrieve procedural and semantic contexts. Stage 3 generates code using the retrieved contexts and static analysis feedback.}
  \label{fig:pipeline}
\end{figure}

ProjAgent is a repository-level code generation system that retrieves procedurally and semantically similar contexts from a repository to provide an LLM with repository-specific guidance for code generation. Given a coding question consisting of an incomplete Python function and its docstring, along with the corresponding repository, ProjAgent generates the completed function. We use Qwen2.5-Coder-14B-Instruct (Qwen2.5) as the backbone model for its strong code-generation capabilities \cite{hui2024qwen2}. Throughout this paper, the function to be completed is referred to as the \textit{target function}, while all other functions in the repository are referred to as \textit{context functions}, which serve as candidate contexts for completing the \textit{target function}.

Figure \ref{fig:pipeline} provides an overview of the ProjAgent pipeline. In the offline preprocessing stage, a subset of repository context functions is decomposed into procedural steps (context steps), and a projection is computed for each step (Section \ref{sec:stage1}). During procedural retrieval, an agentic workflow identifies procedurally similar contexts (Section \ref{sec:stage2a}). Semantic retrieval then retrieves semantically related contexts, including APIs, classes, and utility functions (Section \ref{sec:stage2b}). Finally, the code generation stage completes the target function using the retrieved procedural and semantic contexts and iteratively refines the generated code based on static analysis feedback (Section \ref{sec:stage3}).

\subsection{Offline Preprocessing}\label{sec:stage1}

\noindent\textbf{Function Selection}:\label{sec:func_selection}
The offline preprocessing stage processes repository context functions to construct the projection representations used during procedural retrieval. However, large repositories may contain thousands of context functions (e.g., Astropy contains over 16,000), making preprocessing every function computationally expensive. Moreover, Hu et al.~\cite{hu2026line} showed that most repository files are irrelevant to a given coding question. Therefore, ProjAgent first selects a subset of context functions for preprocessing.

We first identify all Python files in the repository. For each file $f$, we compute two scores: $N_{target}(f)$, the number of target functions in $f$, and $N_f(f)$, the total number of functions in $f$. Files are ranked in descending order by $N_{target}(f)$, with ties broken by $N_f(f)$. This ranking prioritizes files containing benchmark target functions and favors those with richer functional content, consistent with prior work showing that functions within the same file often provide useful contextual information for code generation~\cite{li2024deveval}. We then select the top 20 files, extract all functions, remove the target functions (i.e., the benchmark coding questions), and retain the remaining context functions. The subsequent preprocessing steps operate only on these selected context functions.

\noindent\textbf{Function Decomposition}\label{sec:func_decomp}: After selecting the context functions, we decompose each into a sequence of logical steps, where a logical step represents a distinct operation within the implementation (e.g., input validation). We provide the LLM with the function signature, docstring, and body, and instruct it to identify these logical steps. For each step, the model produces (1) a natural language description capturing its intent and (2) the corresponding code snippet implementing that step. We then validate each generated step by verifying that its code snippet exists in the original function and accurately corresponds to its description, as detailed next.

\noindent\textbf{Step Information Validation}: To verify that a generated context step is grounded in its context function, we propose a three-stage validation process. First, we check whether the function docstring contains the step description. The docstring is extracted using \texttt{ast.get\_docstring} \cite{python_ast}, and both the docstring and step description are normalized following prior work \cite{leusch2005preprocessing, aliero2023systematic} by converting text to lowercase, replacing newlines with spaces, removing punctuation, and collapsing consecutive whitespace. If the normalized step description appears as a substring of the normalized docstring, the step is considered valid. Otherwise, the second stage computes the ROUGE-L score \cite{lin2004rouge} between the normalized texts. Following prior work \cite{joo2025cleanse}, a step is accepted if its ROUGE-L score exceeds 0.7. If this stage also fails, we prompt the LLM with the function body and docstring as the source text, and the step description as the claim, asking it to determine whether the claim is supported by the source text. The model's judgment serves as the final validation decision. The resulting steps are referred to as \textit{valid context steps}.

Next, we validate the code snippet associated with each valid context step. We first verify that every line in the snippet exists in the original function body. If any line is missing, we prompt the LLM with the step description and function body, asking it to re-identify the corresponding code snippet. We then encode the step description and code snippet using \textit{google/embeddinggemma-300m} \cite{vera2025embeddinggemma}, which is trained on both natural language and code and provides an explicit retrieval mode suitable for cross-modal similarity \cite{vera2025embeddinggemma}. The cosine similarity between the two embeddings is computed, and following prior work \cite{chen2026bridging}, the snippet is considered valid if the similarity exceeds 0.75. Only context steps with valid code snippets are retained for the remainder of the pipeline; we refer to them as \textit{grounded context steps}.

\noindent\textbf{Step Projection Construction}: For each grounded context step, we construct a reasoning representation using the reasoning subspace projection technique described in Section \ref{section:background}. Specifically, we prompt the LLM with the step description and the step snippet from the grounded context step, and instruct it to reason about its implementation. We then extract the hidden states from the last layer for all response tokens and project them onto the reasoning subspace, which encodes the model's internal reasoning process. The last layer is chosen because Hu et al. \cite{hu2025harp} showed that shallow-layer vectors are primarily represented in the semantic subspace, whereas deep-layer vectors are more concentrated in the reasoning subspace. We selected response tokens rather than prompt tokens because the prompt shares the common instruction and system message. When the step description of the grounded context step is short, the projection obtained from the prompt tokens contains mostly the model's reasoning about the instruction and the system message. We refer to this set of projections as \textit{P}.

\begin{algorithm}[t]
\caption{Stable $\mu$ and $\text{PC}_1$ Estimation}
\label{alg:stable_debias}
\SetAlgoLined
\KwIn{Ranked repository files, initial projections $\mathcal{P}$}
\KwOut{Stable $\mu$, $\text{pc}_1$}

$(\mu_{\text{old}},\ \text{pc}_{1,\text{old}}) \leftarrow \textsc{Build\_Mu\_$Pc_1$}(\mathcal{P})$\;
$\text{is\_stable} \leftarrow \text{False}$,\ $i \leftarrow 0$\;

\While{\normalfont{not is\_stable}}{
    Select next batch of files\;
    Decompose functions in the new batch into steps\;
    Compute those steps' projections\;
    $\mathcal{P} \leftarrow \mathcal{P}\ \cup\ \text{new projections}$\;
    $(\mu_{\text{curr}},\ \text{PC}_{1,\text{curr}}) \leftarrow \textsc{Build\_Mu\_$Pc_1$}(\mathcal{P})$\;
    
    \If{\normalfont{\textsc{CheckStability}}$(\mu_{\text{old}}, \mu_{\text{curr}})$ \normalfont{and} \normalfont{\textsc{CheckStability}}$(\text{PC}_{1,\text{old}}, \text{PC}_{1,\text{curr}})$}{
        $\text{is\_stable} \leftarrow \text{True}$\;
    }
    \If{all functions have processed}{
        $\text{is\_stable} \leftarrow \text{True}$
    }
    $\mu_{\text{old}} \leftarrow \mu_{\text{curr}}$,\ $\text{PC}_{1,\text{old}} \leftarrow \text{PC}_{1,\text{curr}}$\;
    $i \leftarrow i+1$\;
}
$(\mu,\ \text{PC}_1) \leftarrow \textsc{Build\_Mu\_$PC_1$}(\mathcal{P})$\;
\Return $\mu,\ \text{PC}_1$\;
\end{algorithm}

As mentioned in Section \ref{section:background}, the projections in \textit{P} suffered from the anisotropy phenomenon \cite{ethayarajh2019contextual,timkey2021all}. Therefore, we design an incremental stabilization algorithm to compute a stable $\mu$ and $PC_1$, shown in Algorithm \ref{alg:stable_debias}. The algorithm takes the ranked repository files (obtained from Section \ref{sec:func_selection}) and \textit{P} as input. We then compute the $\mu$ and $PC_1$ from the projection set \textit{P}. After that, we select a new batch of 5 Python files from the ranked repository files (ones not selected in Section \ref{sec:func_selection}). Following the same procedure, we extract all context functions within those 5 files (Line 4), decompose them into context steps (see Section \ref{sec:func_decomp}), and compute their projections (Line 6). We then merge those new projections into \textit{P} (line 7). We compute $\mu_{curr}$ and $PC_{1, curr}$ from P (line 8). Following prior work \cite{yu2026modality}, we check the stability of $\mu$ (line 9) using the relative change formula:

\begin{equation}
\frac{\|\mu_{\text{curr}} - \mu_{\text{old}}\|}{\|\mu_{\text{old}}\|} \leq \tau_{\mu}
\end{equation}
with a threshold of 1e-3 and the stability of $PC_1$ (line 9) using the absolute cosine similarity formula:
\begin{equation}
|\cos(\text{PC}_{1,\text{old}},\ \text{PC}_{1, \text{curr}})| \geq \tau_{\text{PC}_1}
\end{equation}
with a threshold of 0.99. Both the relative change formula and the absolute cosine similarity formula measure how much the current estimates of $\mu$ and $PC_1$ have changed relative to earlier iterations. The while loop stops when either both $\mu$ and $PC_1$ are stable (line 10), or all context functions in the repositories have been processed (line 13). 

After obtaining stable $\mu$ and $PC_1$, we process each projection in \textit{P} to remove the shared dominant direction. Specifically, we apply mean pooling \cite{reimers2019sentence} to the projection as we extract hidden states for all response tokens. Finally, we apply the Equation \ref{eq:debias} in Section \ref{section:background} to the resulting projection.

\subsection{Procedural Context Retrieval}\label{sec:stage2a}
Figure~\ref{fig:pipeline} illustrates the procedural context retrieval pipeline. ProjAgent first decomposes the target function into logical steps (target steps) using its signature and docstring, and computes a projection for each step. It then retrieves procedurally similar contexts through an iterative workflow. Each iteration begins by retrieving an initial set of context steps (Section~\ref{sec:initial_seed}), followed by an agentic search to identify additional candidates (Section~\ref{sec:agentic_seed}). The retrieved contexts are then expanded (Section~\ref{sec:expand}), validated (Section~\ref{sec:verification}), and evaluated using an LLM-based plan confirmation step (Section~\ref{sec:plan_confirmation}) to determine whether sufficient procedural context has been collected for each target step. The process repeats until all target steps have sufficient context or a maximum of five iterations is reached.

\noindent\textbf{Initial Seed Finding}\label{sec:initial_seed}: ProjAgent first identifies an initial set of procedurally similar context steps using projection similarity. We refer to each context step from the processed context functions as a \textit{candidate step}. For each target step, we compute the cosine similarity between its projection and the projection of every candidate step, and rank the candidates accordingly. Candidate steps with projection similarity below 0.75 are discarded. We selected this threshold through empirical calibration on five randomly selected target steps from the \textit{Astropy} repository by evaluating thresholds of 0.7, 0.75, and 0.8, and found that 0.75 best separated procedurally similar candidates from dissimilar ones. To further reduce computational cost, we retain only the top 20 remaining candidates. The retained candidates are then verified using an LLM. Given the target step description, candidate step description, and corresponding code snippet, the model determines whether the candidate performs the same computational operation as the target step, regardless of its application domain. This verification complements projection similarity, which captures coarse procedural similarity but may not fully distinguish semantically equivalent computational patterns. Candidate steps confirmed by the LLM are retained as \textit{initial seeds}.

\noindent\textbf{Agentic Seed Finding}\label{sec:agentic_seed}: While projection similarity retrieves procedurally similar contexts from previously decomposed functions, it cannot discover functions that have not yet been processed. To address this limitation, ProjAgent employs an agentic repository exploration strategy in which an LLM navigates the repository using a set of tools, progressively accumulating knowledge to identify promising context functions.

The agent is initialized with the target step description, target file location, repository structure, and the initial seeds identified in Section~\ref{sec:initial_seed}. At each iteration, it reviews the interaction history, reasons about which function is most likely to contain a procedurally similar step, determines what additional information is needed, and selects the next tool call. To explore the repository, the agent is provided with the \textit{ls}, \textit{execute\_bash}, \textit{read\_func}, \textit{read\_lines}, and \textit{search\_func} tools \cite{zhang2024autocoderover}. Once it identifies a promising function, it invokes \textit{propose\_func}. To encourage sufficient exploration, the agent is instructed to avoid proposing functions prematurely. Table~\ref{tab:agent_tools} summarizes all available tools.

When a function is proposed, it is decomposed following Section~\ref{sec:func_decomp} if it has not been processed previously. We then verify whether any of its steps are procedurally similar to the target step through a four-stage filtering process. First, candidate steps with projection similarity below 0.75 are discarded. Second, candidate steps whose projection similarity to any existing seed exceeds 0.95 are removed to avoid redundancy. Third, the remaining candidates are evaluated by the LLM for procedural similarity. If none are accepted, the agent receives feedback on the reasons for rejection, along with the previously rejected functions, allowing it to refine subsequent exploration. If multiple candidates are accepted, we retain the one with the highest projection similarity while ensuring it remains sufficiently distinct from the existing seeds. The accepted step is then added to the initial seed set.

Repository exploration may require many iterations, causing the conversation history to exceed the model's context window and potentially leading to out-of-memory (OOM) failures \cite{wang2026context}. To mitigate this issue, we adopt an incremental context-compaction strategy inspired by recent context engineering work \cite{bui2026building,zhu2026semaclaw}. The conversation history is partitioned into four components: the system prompt, the initial user message, exploration messages, and proposal messages. The system prompt, initial user message, and proposal messages are preserved in full because they define the task and record the agent's decisions. Only the exploration history is compressed using an LLM-based summarizer \cite{packer2023memgpt}, which preserves all explored files, directories, and referenced context functions while reducing the token footprint.

\begin{table}[!t]
\centering
\small
\begin{tabularx}{0.7\linewidth}{cX}
\hline
\textbf{Tool} & \textbf{Description} \\
\hline
\texttt{ls} & List directory contents \\
\texttt{execute\_bash} & Pattern-based grep search \\
\texttt{read\_func} & Read a specific function \\
\texttt{read\_lines} & Read specific line ranges \\
\texttt{search\_func} & Locate functions by name or class \\
\texttt{propose\_func} & Propose a candidate function as seed \\
\hline
\end{tabularx}
\caption{Tools available to the ProjAgent agentic seed finder (See \ref{sec:agentic_seed}).}
\label{tab:agent_tools}
\end{table}

\noindent\textbf{Seed Expansion}\label{sec:expand}: While the identified seeds in the \textit{initial seeds} are valuable, they alone may not be sufficient to cover all implementation details needed by the model. However, exhaustively running the agentic workflow across the entire repository is too expensive. Therefore, we use projection similarity to efficiently filter a large pool of processed context steps into a small set of promising candidate steps without LLM calls. We then expand the procedurally similar contexts from the remaining promising candidate steps. For each context function, we determine whether it contains a procedurally similar step to the target step by computing the cosine similarity between the projection of each seed step and that of each context step (candidate step) in the context function. After that, we retain only context steps whose cosine similarity exceeds the threshold of 0.65 with at least 2 seeds. We determined these two thresholds empirically (See details in Section \ref{sec:rq2_setup}). 
Requiring agreement across multiple seeds improves robustness, as a candidate step that scores highly on only one seed may be coincidentally similar rather than genuinely procedurally similar. We then collect the remaining context steps and refer to them as the \textit{expand context steps}. Finally, we run an LLM-based verification process on \textit{expand context steps} (covered next). Our rationale is that the remaining context steps still contain false positives, particularly for generic steps such as updating a state value or validating an input, which can superficially resemble unrelated context steps.

\noindent\textbf{Verification}\label{sec:verification}: To control costs, we only run the LLM-based verification process for the top 30 context steps in the \textit{expand context steps}. For each target step, the confirmed seeds are presented to the LLM as calibration examples illustrating what counts as procedurally similar, alongside the problem statement and the context steps to be verified, batched dynamically within a token budget. The LLM independently determines whether each context step is procedurally similar to the target step. We add the verified context steps into the \textit{initial seeds} to form the retrieved context for the target step.

\noindent\textbf{Plan Confirmation}: \label{sec:plan_confirmation} After collecting procedurally similar context, we present the step descriptions of each retrieved context and the target step descriptions to the LLM and instruct it to indicate whether the provided context is sufficient to implement the target step. We designate a target step with sufficient contexts for implementation as the confirmed step, and unconfirmed steps otherwise. For each unconfirmed step, the LLM also produces a refined search query that identifies the missing context required to correctly implement it. We update the unconfirmed step's description by adding the search query at the beginning. We then recalculate its projection using the updated step description. Finally, confirmed steps are saved with their retrieved contexts, while unconfirmed steps proceed to the next iteration, where the updated projection is used to search for new procedurally similar contexts.

\subsection{Semantic Retrieval}\label{sec:stage2b} While procedurally similar contexts can teach the LLM about the target steps' logic, the LLM still needs contexts that are semantically related to the retrieved procedurally similar contexts to produce a correct implementation. Those contexts serve as the ingredients needed to turn the logic into the correct implementation. This is essential because procedurally similar contexts are retrieved based on logical similarity to the target step and may come from anywhere in the repository (as shown in Figure \ref{lst:retrieval_example}), making them potentially inaccessible within the target function's scope.

Starting from the import statements in the Python file containing the target function (target file), ProjAgent finds all accessible files by transitively tracing them using \textit{ast.Import} and \textit{ast.ImportFrom} \cite{python_ast}. ProjAgent then constructs a symbol pool containing functions, classes, and constants from those accessible files, representing the complete set of symbols the target function can directly access at runtime. 

To select semantic contexts that help the model implement the target step, ProjAgent builds an enriched query by combining each target step's description with the step description of each retrieved procedural context. To score each symbol, we compute a score \textit{v} that captures both lexical and semantic relevance between the enriched query \textit{q} and symbol \textit{s}:

\begin{equation}
v = 0.5 \cdot \widehat{\text{BM25}}(q, s) + 0.5 \cdot \widehat{\text{dense}}(q, s)
\label{eq:symbol_score}
\end{equation}
where {\text{BM25}}(q, s) is the BM25 score, representing lexical relevance, and {\text{dense}}(q, s) is the embedding similarity calculated using \textit{sentence-transformers/all-mpnet-base-v2} model \cite{kosenko2024krgp}, representing semantic relevance. The $\hat{\cdot}$ operation enforces the BM25 and dense scores to be in $[0, 1]$. Lastly, we rank the symbols by \textit{v} in descending order and select the top 20 symbols as the semantic contexts for each target step.

\subsection{Code Generation With Feedback}\label{sec:stage3}

After retrieving procedural and semantic contexts, ProjAgent formats them separately to maximize target-step coverage within the prompt budget. Each procedural context includes a natural-language step description and its corresponding code snippet, while each semantic context includes the retrieved code symbols. For each target step, ProjAgent prioritizes confirmed procedural contexts when available and otherwise uses the best unconfirmed context. The same strategy is applied to semantic contexts. All contexts are explicitly labeled as confirmed or unconfirmed to indicate their reliability. Once each target step is covered by at least one procedural and one semantic context, any remaining prompt budget is filled with additional confirmed contexts, prioritizing procedural contexts before semantic ones. ProjAgent then constructs the generation prompt using the problem statement, target-step descriptions, and formatted procedural and semantic contexts, and instructs the LLM to generate the implementation.

ProjAgent further refines the generated code using static-analysis feedback. It first performs an AST-based syntax check \cite{python_ast}; if a syntax error is found, the error message is returned to the LLM. Otherwise, ProjAgent applies four semantic consistency checks inspired by prior work \cite{chen2024teaching}: (1) method call checking verifies that invoked methods are accessible from the current class hierarchy and called with the correct number of arguments; (2) field access checking verifies that referenced fields are defined in the current class hierarchy; (3) variable-method checking infers the receiver type at a call site and verifies that the invoked method exists in the corresponding type hierarchy; and (4) standalone function call checking verifies that called functions resolve to accessible functions and are invoked with the correct number of arguments.

Because Python's dynamic typing prevents complete static resolution, the feedback loop is conservative: an issue is reported only when it can be confirmed. ProjAgent then constructs a refinement prompt containing the previous implementation, problem statement, target-step descriptions, formatted contexts, and static-analysis feedback, and asks the LLM to revise the code. This loop continues until no feedback is produced or the maximum number of iterations (10) is reached.

\section{Experimental Setup}
\label{section:exp_setup}
\subsection{Benchmark and Metric}
To measure the effectiveness of ProjAgent (RQ1) and its components (RQ3), we evaluated ProjAgent on REPOCOD \cite{liang2025can}, a benchmark for repository-level code generation consisting of 980 problems drawn from 11 real-world repositories. Each coding problem requires generating a complete target function given the function signature and docstring. We evaluated the generated code using the test cases provided in the REPOCOD dataset and reported the Pass@1 metric. To control LLM's non-determinism, we used greedy decoding \cite{song-etal-2025-good} by setting temperature 0 and a maximum of 4096 output tokens for code generation.

\subsection{Baselines} We compare ProjAgent against four baselines: dense retrieval (DENSE), sparse retrieval (SPARSE), same-file retrieval (SAME\_FILE), and SpecAgent \cite{ma2025specagent}. DENSE, SPARSE, and SAME\_FILE were selected because they represent complementary repository retrieval strategies, and we follow their implementation as described in REPOCOD \cite{liang2025can}. For DENSE, we use the \textit{sentence-transformers/all-mpnet-base-v2} embedding model \cite{kosenko2024krgp}. We also compare against SpecAgent, the current state-of-the-art repository-level code generation system \cite{ma2025specagent}. As its replication package is unavailable, we have reimplemented the approach using the methodology described in the paper and the prompts provided in its appendix. To ensure a fair comparison, all baselines use Qwen2.5-Coder-14B-Instruct as the backbone model.

\subsection{Projection Effectiveness Setup}\label{sec:rq2_setup}

To evaluate how effectively projection similarity identifies procedurally related contexts (RQ2), we designed a three-step process: (1) construct a labeled dataset, (2) evaluate different projection similarity configurations, and (3) compare their ability to distinguish procedurally similar from dissimilar contexts.

\noindent\textbf{Dataset Construction}: We constructed the dataset using all target functions from the \textit{Astropy} repository in REPOCOD \cite{liang2025can}. For each target step, we executed the initial seed finding, agentic seed finding, and seed expansion stages (Section \ref{sec:stage2a}). For every target-context step pair, we recorded the projection similarity scores between the context step and each expanded seed of the target step. The resulting dataset maps each (target step, context step) pair to its corresponding list of similarity scores. Overall, it contains 9,598 target-context step pairs spanning all 85 target functions and every context function in the repository.

\noindent\textbf{Dataset Labeling}: We labeled each of the 9,598 pairs using Claude Sonnet 4.6 \cite{anthropic2025claude}. For each pair, Claude received the target and context step descriptions, along with the context step snippet, and determined whether the context step was procedurally similar to the target step. To clarify the notion of procedural similarity, we employed contrastive in-context learning \cite{gao2024customizing} using one manually selected positive and one negative example from the \textit{Seaborn} repository.

To validate the automated labels, two authors independently annotated a statistically significant random sample of 370 pairs (95\% confidence level, 5\% margin of error). Each author assessed whether the context step exhibited a computational pattern similar to that of the target step, based on the step descriptions and the code snippet. Disagreements were resolved through discussion. The resulting inter-rater agreement was $\kappa=0.86$, indicating almost perfect agreement \cite{landis1977measurement}. Comparing the consensus labels with Claude's predictions yielded $\kappa=0.82$, demonstrating strong agreement.

\noindent\textbf{Evaluation}: For each projection similarity configuration, defined by the similarity threshold $\tau$ and agreement threshold $k$, we partitioned the dataset into a \textit{promoted group} and a \textit{leftover group}. A target-context pair was assigned to the promoted group if at least $k$ projection similarity scores exceeded $\tau$; otherwise, it was assigned to the leftover group. We evaluated six $(\tau,k)$ configurations and report precision, recall, and F1 score for both groups using the manually validated labels as ground truth. We restricted the analysis to a single repository for two reasons: manually validating all candidate pairs across the 11 REPOCOD repositories is infeasible, and selecting in-context examples from the evaluation repository could introduce information leakage that biases Claude's judgments.

\noindent\textbf{Experimental Scope:} Due to the computational cost of repository-level code generation, all experiments were conducted using a single backbone model (Qwen2.5-Coder-14B-Instruct). A complete evaluation of one system requires generating and executing code for all 980 REPOCOD tasks across 11 repositories. Beyond the main evaluation, this work includes multiple baselines, six projection-similarity configurations for RQ2, and four system variants for the RQ3 ablation study, resulting in several thousand end-to-end code generation and test-execution runs. Repeating the entire experimental pipeline across multiple backbone models would increase the computational cost proportionally while substantially extending the experimental time. We therefore focus on a single, strong open-source code model to enable a comprehensive evaluation of the proposed retrieval methodology under a consistent experimental setting.

\subsection{Ablation Setup for RQ3}
To understand the effectiveness of each retrieval component, we conducted an ablation study on all 85 coding problems from the \textit{Astropy} repository in REPOCOD. We evaluated three variations: \textbf{w/o procedural}, which removes the contexts retrieved from the procedurally similar signal; \textbf{w/o semantic}, which removes the contexts retrieved from the semantic retrieval signal; and \textbf{w/o feedback loop}, which uses contexts collected from both procedurally similar and semantic retrieval signals, but disables the static analysis feedback loop. 
In each variant, only the specified component was removed, while all others remained identical to those in the full pipeline. We report the Pass@1 score for each variant.

\section{Result}
\label{section:result}

\noindent\textbf{Answer to RQ1}: Table \ref{tab:main_results} summarizes the code generation performance of ProjAgent and the baselines on the REPOCOD benchmark. ProjAgent achieves the highest Pass@1 among all methods. Compared to SPARSE, DENSE, and SAME\_FILE, ProjAgent improves Pass@1 by 12.31\%, while outperforming SpecAgent by 6.62\%.

We posit that these gains stem from the quality of the retrieved procedural and semantic contexts. However, because repository exploration is constrained by a fixed search budget, the agent may fail to discover highly relevant procedural contexts. To assess the impact of search coverage, we conducted an additional experiment on the \textit{Astropy} repository. Instead of decomposing context functions on demand, we decomposed every context function offline and applied the initial seed finding, seed expansion, verification, and plan confirmation pipeline (Section \ref{sec:stage2a}) to all of them. Under this full-search setting, ProjAgent achieved a Pass@1 of 25.12\%, compared to 21.16\% under the budget-constrained agentic search, representing a 3.96\% improvement. These results indicate that search coverage is a key factor in ProjAgent's effectiveness and suggest that future work should focus on improving repository exploration under fixed search budgets.

\begin{tcolorbox}[colback=gray!10, colframe=black, boxrule=0.5pt, arc=2pt, left=4pt, right=4pt, top=4pt, bottom=4pt]
\textbf{RQ1 Key Takeaway:} Procedural retrieval substantially (\textbf{6.62\%}) improves repository-level code generation. While ProjAgent outperforms all baselines, additional gains are observed as repository search coverage increases, suggesting that efficient exploration of large repositories is a primary bottleneck for further performance improvements.
\end{tcolorbox}

\begin{table}[t]
\centering
\strut
\begin{tabularx}{0.4\linewidth}{Xc}
\toprule
\textbf{Method} & \textbf{Pass@1}\\
\midrule
Sparse             & 26.58\%\\
Dense              & 28.83\%\\
Same\_File         & 14.98\%\\
SpecAgent          & 34.52\%\\
\midrule
\textbf{ProjAgent (Ours)} & \textbf{41.14\%}\\
\bottomrule
\end{tabularx}
\caption{Pass@1 (\%) on REPOCOD for all methods using Qwen2.5-Coder-14B-Instruct.}
\label{tab:main_results}
\end{table}

\begin{table*}[t]
\centering
\begin{tabularx}{\linewidth}{l X c c c c c c c c}
\toprule
& & \multicolumn{4}{c}{\textbf{Promoted}} & \multicolumn{4}{c}{\textbf{Leftover}} \\
\cmidrule(lr){3-6} \cmidrule(lr){7-10}
\textbf{Cfg} & \textbf{($\tau$, $k$)} & \textbf{\#Pairs} & \textbf{Prec.} & \textbf{Recall} & \textbf{F1} & \textbf{\#Pairs} & \textbf{Prec.} & \textbf{Recall} & \textbf{F1} \\
\midrule
A & (0.60, 2) & 2,036 & 3.6\% & 36.7\% & 0.065 & 7,562 & 98.3\% & 79.1\% & 0.877 \\
\textbf{B} & \textbf{(0.65, 2)} & \textbf{997} & \textbf{5.4\%} & \textbf{27.1\%} & \textbf{0.090} & \textbf{8,601} & \textbf{98.3\%} & \textbf{90.0\%} & \textbf{0.940} \\
C & (0.65, 3) & 677   & 5.8\% & 19.6\% & 0.089 & 8,921 & 98.2\% & 93.2\% & 0.956 \\
D & (0.70, 2) & 410   & 6.6\% & 13.6\% & 0.089 & 9,188 & 98.1\% & 95.9\% & 0.970 \\
E & (0.70, 3) & 294   & 6.1\% & 9.0\%  & 0.073 & 9,304 & 98.1\% & 97.1\% & 0.976 \\
F & (0.80, 2) & 45    & 6.7\% & 1.5\%  & 0.025 & 9,553 & 97.9\% & 99.6\% & 0.987 \\
\bottomrule
\end{tabularx}
\caption{Precision, recall, and F1 score of the promoted and leftover groups across six threshold configurations. Config B (bolded) achieves the highest Promoted-F1, indicating the best precision-recall balance.}
\label{tab:rq2}
\end{table*}

\noindent\textbf{Answer to RQ2}: Table \ref{tab:rq2} reports the number of pairs per group (promoted or leftover), the precision, recall, and F1 score of each group across the six configurations, with the best configuration (Configuration B) shown in bold. From the precision, recall, and F1 score for the \textit{leftover group}, we observed that these scores remained consistently high across all configurations. This indicates that the projection similarity-based method reliably excludes procedurally dissimilar context steps regardless of the chosen configuration. On the other hand, the precision, recall, and F1 score of the promoted group across all six configurations are low. We suspect these low scores are due to two factors. First, not every target step has a procedurally similar context step in the repository, because some target steps implement highly specific logic unique to their functions. To validate this, we took the retrieved procedurally similar contexts for each coding question across all repositories and calculated the ratio of query steps that have at least one procedurally similar context. We found that only 35.1\% of query steps have at least one genuine procedurally similar context step. Second, many target and context steps describe generic operations, such as validating input and checking input. These steps share similar computational patterns, but only at a level of abstraction that is too coarse to be procedurally similar. This leads to false positives in the promoted group, directly lowering its precision. We further observed that precision increases monotonically as the thresholds $\tau$ and $k$ become stricter, from 3.6\% (configuration A) to 6.7\% (configuration F), while recall drops sharply from 36.7\% (configuration A) to 1.5\% (configuration F). This indicates that the stricter configuration discards the vast majority of procedurally similar context steps. Since the precision, recall, and F1 scores of the \textit{leftover group} remain consistently high, we select the best configuration based on the F1 score of the \textit{promoted group} and found that configuration B ($\tau$ of 0.65 and agreement $k$ of 2) performs the best. The low precision, recall, and F1 score in the \textit{promoted group} also justifies the verification step described in Section \ref{sec:verification}, as it filters out the false positives in the promoted group and improves the quality of the retrieved procedurally similar contexts. 

Overall, these results suggest that projection similarity is most effective as a \emph{candidate generation} mechanism rather than a standalone procedural similarity classifier. While the promoted group contains many false positives, the consistently high precision of the leftover group indicates that projection similarity effectively eliminates a large portion of procedurally unrelated context steps from consideration. This substantially reduces the search space while retaining most candidate procedural matches, allowing the subsequent verification stage to focus on a much smaller set of potentially relevant contexts. The complementary strengths of projection similarity and verification, therefore, motivate the two-stage retrieval design adopted by ProjAgent.

\begin{tcolorbox}[colback=gray!10, colframe=black, boxrule=0.5pt, arc=2pt, left=4pt, right=4pt, top=4pt, bottom=4pt]
\textbf{RQ2 Key Takeaway:} The projection similarity-based method reliably filters out procedurally dissimilar context steps from a large pool of context functions (leftover precision $\geq$ 97.9\% across all configurations). \textbf{Configuration B} ($\tau$=0.65, $k\geq2$) achieves the best F1 score and is adopted as the default configuration of ProjAgent.
\end{tcolorbox}

\noindent\textbf{Answer to RQ3}: Table \ref{tab:rq3_result} presents the results of the ablation study. Removing procedurally or semantically similar contexts results in a substantial drop in Pass@1, confirming that both types of contexts are critical to ProjAgent's performance. When dropping the procedurally similar contexts, the performance drops from 41.14\% to 25.76\%. On the other hand, when semantically similar contexts are dropped, performance drops from 41.14\% to 32.29\%. Comparing \textit{wo/procedural} with \textit{wo/semantic}, removing procedurally similar contexts results in greater performance degradation for ProjAgent.

We attribute this difference to the distinct roles of the two retrieval mechanisms. Procedural retrieval provides examples of implementation procedures that guide how the target functionality should be implemented, whereas semantic retrieval supplies repository-specific symbols, such as functions, classes, and constants, that enable those procedures to be translated into executable code. Together, these retrieval signals provide both procedural guidance and repository-specific knowledge, explaining why combining them yields substantially better performance than using either retrieval signal alone.

\begin{table}[t]
\centering
\begin{tabularx}{0.5\linewidth}{Xc}
\toprule
\textbf{Method} & \textbf{Pass@1}\\
\midrule
ProjAgent (All Components) & \textbf{41.14\%}\\
\midrule
wo/procedural      & 25.76\%\\
wo/semantic        & 32.29\%\\
wo/feedback loop   & 40.29\%\\
\bottomrule
\end{tabularx}
\caption{Ablation on Different Components of ProjAgent}
\label{tab:rq3_result}
\end{table}

In contrast, removing the static-analysis feedback loop produces only a modest decrease in performance. We attribute this to the conservative design of the feedback mechanism, which reports only errors that can be determined with high confidence. Consequently, many ambiguous issues, such as those involving dynamically inferred argument types, are intentionally ignored to avoid introducing incorrect feedback. While this design minimizes false corrections, it also limits opportunities for iterative refinement. Future work could investigate richer feedback mechanisms, including runtime execution feedback and lightweight type inference, to improve the effectiveness of the refinement stage.

\begin{tcolorbox}[colback=gray!10, colframe=black, boxrule=0.5pt, arc=2pt, left=4pt, right=4pt, top=4pt, bottom=4pt]
\textbf{RQ3 Key Takeaway:} %
Procedural and semantic retrieval provide complementary information for repository-level code generation, and removing either substantially reduces Pass@1. Within the current ProjAgent pipeline, removing procedural retrieval results in the larger performance degradation, while the conservative static-analysis feedback loop provides a modest but measurable improvement.
\end{tcolorbox}

\section{Threat To Validity}
\label{section:ttv}
\noindent\textbf{External Validity:} We evaluated ProjAgent on REPOCOD, a repository-level code generation benchmark consisting exclusively of Python repositories. Consequently, our findings may not generalize to repositories written in other programming languages or to other repository-level code generation benchmarks. In addition, ProjAgent was implemented using Qwen2.5-Coder-14B-Instruct as the backbone model. The effectiveness of procedural retrieval may vary across LLMs with different hidden-state representations, potentially affecting the quality of reasoning-subspace projections. We limited our evaluation to a single backbone model because repository-level code generation is computationally expensive: a single evaluation requires generating and executing code for all 980 tasks in REPOCOD. Our study also includes four baselines, six projection-similarity configurations for RQ2, and four variants for the RQ3 ablation study. Extending this evaluation to multiple backbone models would increase the computational cost proportionally. Finally, the ablation study for RQ3 was conducted on a single repository (\textit{Astropy}); therefore, the relative contribution of individual retrieval components may differ for repositories with different structures, domains, or dependency characteristics. In addition, our experiments used a fixed retrieval budget and context window. The effectiveness of procedural retrieval may differ under alternative retrieval budgets or models with substantially larger context windows.

\noindent\textbf{Internal Validity:} In RQ2, we used Claude as an LLM judge to label candidate procedural relationships because manually annotating the large number of (target function step, context function step) pairs required for evaluation was prohibitively time-consuming. Although LLM judges may produce incorrect labels, we mitigated this threat by having two researchers independently manually inspect a statistically significant random sample of 370 pairs and verify the assigned labels. In addition, ProjAgent's performance is sensitive to several configuration choices, including the projection-similarity threshold, seed-agreement count, and the number of iterations used for agentic seed discovery. While we conducted ablation studies to assess the contributions of individual components, we did not exhaustively explore all possible parameter configurations due to the substantial computational cost of evaluating repository-level code generation systems. Consequently, different parameter settings may lead to different performance outcomes.

\section{Conclusion}
\label{section:conclusion}

In this paper, we introduced \textit{procedural similarity} as a complementary retrieval signal for repository-level code generation and presented \textsc{ProjAgent}, a retrieval-augmented generation system that integrates procedural, lexical, and semantic retrieval within an agentic workflow, together with a static-analysis feedback loop. Unlike existing retrieval approaches that primarily rely on lexical, semantic, or structural similarity, ProjAgent identifies repository functions that implement similar computational procedures, thereby providing LLMs with context useful for solving the target programming task.

Our evaluation on REPOCOD demonstrates that ProjAgent achieves state-of-the-art repository-level code generation performance, outperforming existing baselines. We further show that reasoning-subspace projections provide an effective representation for identifying procedurally related functions and that combining procedural retrieval with conventional retrieval signals consistently improves generation performance. Ablation studies further confirm that each component of the retrieval pipeline contributes to the system's overall effectiveness.

More broadly, our results suggest that retrieval quality depends not only on identifying code that is lexically or semantically similar, but also on retrieving code that follows similar computational procedures. We hope this work motivates future research on behavior-aware retrieval methods for software engineering tasks and on richer representations of procedural knowledge in large language models.

\bibliographystyle{ACM-Reference-Format}
\bibliography{refs}

\end{document}